\documentclass[aps,prl,floatfix,twocolumn,superscriptaddress]{revtex4}

\usepackage{graphicx}
\usepackage{graphics}
\usepackage{amssymb}
\usepackage{amsmath}
\usepackage{epsfig}
\usepackage{latexsym}
\usepackage{color}
\usepackage{subfigure}
\usepackage{array}
\usepackage{bm}

\newcommand{\CNOT}[0]{\mathrm{CNOT}}
\newcommand{\Tr}[1]{\mathrm{\mathrm{Tr} \left({#1}\right)}}

\newcommand{\X}[0]{\sigma_{X}}
\newcommand{\Y}[0]{\sigma_{Y}}
\newcommand{\Z}[0]{\sigma_{Z}}

\newcommand{\Xp}[1]{X_\pi^{(#1)}}

\newcommand{\Zp}[1]{Z_\pi^{(#1)}}

\newcommand{\Ry}[2]{Y_{#1}^{(#2)}}
\newcommand{\Rz}[2]{Z_{#1}^{(#2)}}

\begin{document}

\title{Robust CNOT gates from almost any interaction}

\author{Charles D. Hill}
\email{Charles.Hill@liverpool.ac.uk} \affiliation{Department of
Electrical Engineering and Electronics, University of Liverpool,
Brownlow Hill, L69, Liverpool, United Kingdom.}


\begin{abstract}
There are many cases where the interaction between two qubits is not
precisely known, but single qubit operations are available. In this
paper we show how, regardless of an incomplete knowledge of the
strength or form of the interaction between two qubits, it is often
possible to construct a CNOT gate which has arbitrarily high
fidelity. In particular, we show that oscillations in the strength
of the exchange interaction in solid state Si and Ge structures are
correctable.
\end{abstract}

\maketitle



Any realistic quantum computer has errors. Principally these errors
come in two varieties: random decoherence and systematic errors.
Systematic errors can arise from imperfections and inhomogeneities
in the construction or implementation of demanding experiments. Both
systematic errors and errors due to decoherence may be corrected,
although it is considerably easier to correct systematic errors.

A pertinent example of systematic error is the strength of the
exchange interaction oscillation in solid state silicon based
architectures \cite{KHD+03,KHS02, WHP+03,WHK+04, WH05, Ket06}. The
six conduction-band minima in silicon generate inter-valley
electronic interference. Discouragingly this causes oscillation in
magnitude of the exchange splitting between two neighboring donors.
The strength of the interaction between qubits therefore sensitively
depends on the exact positioning of donors. In this paper we
demonstrate that, in principle, such a systematic error in the
strength or form of interaction between two qubits \emph{is}
correctable.

Systematic errors may be corrected using composite pulses, in which
a single operation is replaced by several imperfect pulses in such a
way that systematic errors in each pulse cancel each other. Freeman
\cite{Fre97} and Levitt's review \cite{Lev86} and the references
therein provide an excellent introduction. More recently Jones
\cite{Jon03} notes that single qubit composite pulses can modified
to apply to the Ising interaction. In particular he presents a two
qubit pulse sequence based on those by Wimperis \cite{Wim84} for the
construction of a CNOT gate in NMR.

This paper applies to any architecture with the ability to apply
single qubit rotations and a coupling between the two qubits.
Therefore many leading quantum computing architectures - including
solid state architectures - can, in principle, correct for an
unknown coupling between qubits. This addresses a common problem
across many architectures, where composite pulses have begun to be
applied (for example in ion traps \cite{RHR+04, BCS+04, GRL+04} and
Josephson Junctions \cite{CIA+04}). As an example, we explicitly
consider electron spin in the Kane architecture \cite{Kan98}.

Using the method presented here, it is not necessary to know either
the \emph{strength} or the \emph{form} of the coupling. We will not
assume that the error is in the strength of the interaction alone. In
fact, we will demonstrate that it is possible to create a high
fidelity CNOT gate from a largely random Hamiltonian.

A key benefit of composite pulses is that the error does not need to
be perfectly characterised. Characterising the strength and form of
the interaction to a high degree of accuracy is a challenging
task. Even with an accurate characterisation of the Hamiltonian, the
pulse sequences given in this paper outperform a naive implementation
of the CNOT gate. Although we never \emph{learn} the exact
Hamiltonian, we arrange that systematic errors cancel themselves.

The composite CNOT gate construction follows the following steps:
\begin{enumerate}
\item Isolate a single term: In this step, a single coupling terms is
  isolated from the interaction Hamiltonian.

\item Create a composite control sign gate: In this step, pulses
  adapted from NMR correct for systematic errors in the strength of
  the coupling.

\item Finally, apply single qubit unitaries.
\end{enumerate}

A completely general two-qubit Hamiltonian may be expanded in the
Pauli basis as
\begin{equation}
H_2 = \sum_{i,j = \{I,X,Y,Z\}} J_{ij} \sigma_i \sigma_j.
\end{equation}
where $\sigma_i$ are the Pauli matrices, and as throughout the
paper, the tensor product is implied. This Hamiltonian includes both
coupling between the qubits and single qubit terms. The coupling
energies between the qubits are given by the constants $J_{ij}$
($i\ne I, j\ne I$). We do not assume that we know either the
strength of the single qubit terms, or the coupling terms. There
will be a coupling energy which we believe is greatest. Without loss
of generality, let us assume that this term is $J_{ZZ}$. Any single
two qubit term is sufficient.


It is well known that it is possible to isolate a particular term of
the interaction using a technique called \emph{term isolation}
\cite{BDN+04}. In our case, it is possible to isolate the $J_{ZZ}$
term. Consider the pulse sequence
\begin{eqnarray}
Q(t) &=& \Zp{1} \Zp{2} \ V_{t/4} \ \Zp{1} \ V_{t/4} \ \Zp{1} \Zp{2} \
V_{t/4} \ \Zp{1} \ V_{t/4}, \label{eqn:refocus1}\\
V_t &=& \Xp{1} \Xp{2} \ \exp\left(i H_2 \frac{t}{2}\right)
\Xp{1} \Xp{2} \exp\left(i H_2 \frac{t}{2}\right)
\end{eqnarray}
Here, as throughout this paper, a single qubit rotation of an angle
$\theta$ around the z-axis of the $i^{th}$ qubit is denoted by
\begin{equation}
\Rz{\theta}{i} = \exp\left(i \frac{\theta}{2} \sigma_{Z}\right),
\end{equation}
and similarly for rotations around the x and y-axes. This pulse
sequence isolates a single coupling term:
\begin{equation}
Q(t) \approx \exp(i J_{ZZ}t \ \Z\Z). \label{eqn:Q}\\
\end{equation}
Eq. (\ref{eqn:Q}) is only correct to first order, because not all
terms in the Hamiltonian, $H_2$, commute. However, it may be made
arbitrarily accurate by applying the pulses, $\Xp{1}\Xp{2}$,
$\Zp{1}\Zp{2}$ and $\Zp{1}$, $k$ times more frequently:
\begin{equation}
Q_k(t) = Q(t/k)^k
\end{equation}

Term isolation is not uniformly valid. If there is no coupling of the
specified type (that is $J_{ZZ} =0$) then the qubits will be decoupled
by the pulse sequence, and no term isolated. Also, to perform term
isolation it is necessary that the single qubit rotations are
implemented much faster than the typical timescale of the coupling
between qubits. This requires either fast single qubit rotations, or
the ability to turn the interaction between qubits on and off.

If interaction Hamiltonian is \emph{known} to have a simpler form,
then a single coupling term may be isolated more simply and
effectively. For the Heisenberg interaction:
\begin{equation}
H_H = J(\X \X + \Y\Y + \Z\Z),
\end{equation}
all terms commute, and therefore $J_{ZZ}$ be isolated using just two
steps:
\begin{equation}
\exp(i J_{ZZ}t \Z\Z) = \Zp{1} \ \exp \left(i H_H t \right) \ \Zp{1} \
\exp \left(i H_H t \right). \label{eqn:heisenberg} \\
\end{equation}
Eq. \eqref{eqn:heisenberg} is exact, and would only need to be
carried out once. For many systems, such as the nuclei and electron
spins in the Kane architecture, or quantum dots, this much simpler
form of term isolation may be used.

This completes the first step: To isolate a coupling single term.
For a completely general two qubit Hamiltonian, it is always
possible to isolate a single coupling term. The strength of this
term remains unknown, but as this paper now shows, systematic errors
in the strength $J_{ZZ}$ can be corrected.


The exact coupling strength, $J_{ZZ}$ is not known. In general we
will predict a certain value, $J_P$. Unless the gate is perfectly
characterised, we will make some fractional error, $\Delta$, defined
as $J_{ZZ} = (1+\Delta){J_{P}}$. Therefore, when we attempt to
create the gate
\begin{equation}
\theta_0 = \exp\left(i \frac{\theta}{2} \Z\Z\right),
\end{equation}
by setting $t=\frac{\theta}{J_P}$ we will systematically over-rotate
or under-rotate, \emph{actually} creating the gate
\begin{equation}
\theta^{[1]}_0 = \left(\theta(1+\Delta)\right)_0 \approx Q(t).
\end{equation}

Jones \cite{Jon03} notes that single qubit composite pulses can be
modified to apply to the Ising interaction. In particular a two
qubit pulse sequence based on BB1 \cite{Wim84} is presented. The
symmeterized version of the pulse is
\begin{equation}
\theta_0^{[2]} = (\theta/2)^{[1]}_0 \ \pi^{[1]}_{\phi} \
2\pi^{[1]}_{3\phi} \ \pi^{[1]}_{\phi} \ (\theta/2)^{[1]}_0,
\label{eqn:comp1}
\end{equation}
where this pulse is made up of imperfect gates,
\begin{equation}
\theta_\phi = \Ry{\phi}{2} \ \theta_0 \ \Ry{-\phi}{2}
\end{equation}
and in order to cancel first and second order terms, $\phi =
\arccos\left(-\frac{\theta}{4\pi}\right)$.

An alternative pulse which gives the same increase in fidelity when
the uncertainty in $J_{ZZ}$ is the only source of error, but
which allows us to refocus an additional time, is given by
\begin{equation}
\theta_0^{[2]} = (\theta/2)_0^{[1]} \ \frac{\pi}{2}^{[1]}_\phi \
\frac{\pi}{2}^{[1]}_{-\phi} \ \Zp{2} \ \frac{\pi}{2}^{[1]}_\phi \
\frac{\pi}{2}^{[1]}_{-\phi} \ (\theta/2)^{[1]}_0 \ \Zp{2}.
\label{eqn:comp2}
\end{equation}

Pulse schemes on a single qubit may be made arbitrarily accurate
\cite{BHC04}. This is also true of two qubit pulses. One
straight-forward way to do this is to feed the pulse back into
itself. If we implement the pulse sequence,
\begin{equation}
\theta_0^{[2*]} = \left(\Xp{2} \ \Xp{2} \ \Zp{2} \
(\theta/16)^{[2]}_0 \ \Zp{2}\ (\theta/16)^ {[2]}_0 \right)^{8},
\label{eqn:comp3}
\end{equation}
then by feeding this pulse back into the right hand side of Eq.
\eqref{eqn:comp2}, we obtain a pulse which is correct to higher
order. In principle there is no limit to the order which is
achievable.

The average fidelity, for the purposes of this paper, is defined as
\begin{equation}
F^2 = \frac{|\Tr{U_I^{\dagger}U}|}{\Tr{U^{\dagger}U}}
\end{equation}
where $U_I$ is the actually implemented operation, and $U$ is the
intended rotation.

\begin{figure}
\begin{center}
\includegraphics[height=\columnwidth, angle=-90]{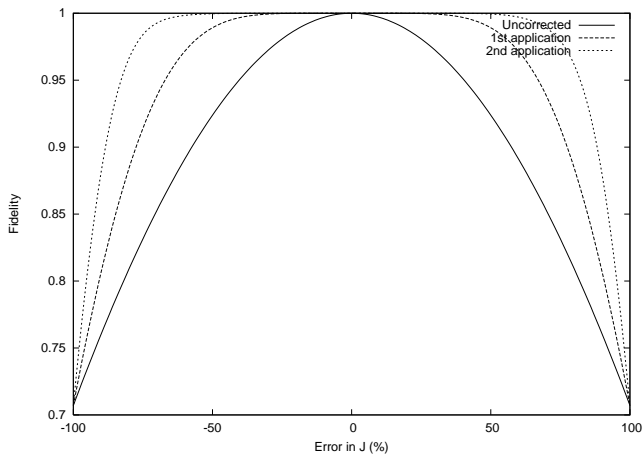}
\caption{This plot shows the fidelity of several methods of creating a
CNOT gate, with a systematic error in the strength of the
coupling.} \label{fig:graph}
\end{center}
\end{figure}

Using each of the three pulse sequences, we attempted to create the
entangling component of the CNOT gate,
$\left(\frac{\pi}{2}\right)_0$. Fig. \ref{fig:graph} shows the
fidelity each pulse sequence, plotted against the error, $\Delta$,
in the strength of the interaction. The solid line shows the
fidelity without any correction. The first dotted line shows the
fidelity of the composite pulse described in Eq. (\ref{eqn:comp1})
or Eq. (\ref{eqn:comp2}). The composite pulse provides an
improvement over the fidelity of the uncorrected pulse for $J_{ZZ}(1
\pm 100)\%$. The higher order pulse described using Eq.
(\ref{eqn:comp3}) is shown as the second dotted line. It shows an
improvement over both the uncorrected pulse and the first composite
pulse between $\Delta=-100\%$ and $\Delta=100\%$.

If we wish to have an error of $1 \times 10^{-4}$ then without
correction we require a $\Delta$ of less than 1\%. For the composite
pulse scheme described by Eq. (\ref{eqn:comp1}) or
Eq. (\ref{eqn:comp2}), we may tolerate an error, $\Delta$ of
approximately 22\%. In the higher order composite pulse described
using Eq. (\ref{eqn:comp3}), an error $\Delta$ approximately 41\%
still achieves a fidelity of $99.99\%$.

This concludes the second step. A systematic error in the interaction
strength, $\Delta$ may be corrected using two-qubit extensions of well
known composite pulses. These pulses may be made arbitrarily accurate
by concatenation.

For the final step, we simply note that a CNOT gate may be written as
\begin{equation}
\CNOT = H^{(2)} \ \Rz{\frac{\pi}{2}}{1} \Rz{\frac{\pi}{2}}{2}
\exp\left(i \frac{\pi}{4}\ \Z\Z\right) \ H^{(2)}.
\end{equation}

A robust CNOT gate may be constructed applying all three steps. The
first step isolates the $\Z\Z$ term, regardless of the form of the
Hamiltonian. The second step corrects for any error in the strength of
this term, and finally the third step applies single qubit unitaries
to complete the \emph{robust} CNOT. Using this robust CNOT gate, we
now describe two examples.

One of the current concerns about the viability of the construction
of an exchange based solid state quantum computer is oscillations in
the strength of the exchange interaction \cite{KHS02, KHDS02,
Ket06}. For an arbitrarily placed donor, the strength of the
exchange interaction is unknown. Even the variation of the donor's
position by a single lattice site can change the strength of the
exchange interaction dramatically. The placement introduces an
unknown systematic error in the strength of the exchange
interaction. Fortunately, that is \emph{exactly} the type of error
which is corrected in this paper. It does not matter that we do not
know the strength of the interaction, or that the exchange
interaction may differ from site to site.

For the Kane quantum computer the single-qubit Hamiltonian is given
by
\begin{equation}
H_{Q}=\mu_{B}B\sigma_{e}^{z}-g_{n}\mu_{n}B\sigma_{n}^{z}+A(V_{A})
{\sigma}_{e}\cdot{\sigma}_{n},\label{eqnE:HQ}
\end{equation}
where $B$ is the strength of the constant magnetic field,
$\sigma^{z}$ is the Pauli Z matrix with subscripts $e$ referring to
electrons and $n$ referring to the nucleus and $A(V_A)$ is the
strength of the hyperfine interaction. This allows single qubit
operation of the computer using either the nuclear spin as a qubit
\cite{Kan98} or electron spin \cite{Hil05}. The exchange coupling
between electrons whose strengths, $J_i$, can vary considerably,
leads to the Hamiltonian
\begin{equation}
H = \sum_i J_i \sigma_i \cdot \sigma_{i+1} + H_Q^{(i)}.
\end{equation}
This Hamiltonian allows for the single qubit operation of the
computer, but it is not immediately clear if two qubit operations
can be implemented while the strengths of the exchange interactions
remain unknown.

We now consider the pulses and overhead required to create a
composite CNOT gate using electron spin. We use some simplifying
assumptions. We assume that any single qubit rotation by $\pi$
requires 40ns, as does the Hadamard gate. We also assume that two
qubit rotations by $\pi/8$ require only 1ns. The typical `square
root of swap' CNOT gate requires a total of 6 single qubit gates,
and 2 two-qubit gates. The total time required for a CNOT gate is
approximately $140 ns$. Jones' pulse sequence requires more pulses.
A total of 16 single qubit gates, and 8 two qubit gates are
required. A CNOT gate requires approximately $460 ns$ to complete.
The composite pulse is short compared to the comparatively long
decoherence time of donors in Si.  $1 \times 10^{5}$ operations may
be performed during the $60 ms$ dephasing time measured in bulk Si.
It therefore falls below the fault tolerant threshold \cite{Got97}.

Assuming perfect single qubit gates, the fidelities for these two
sequences exactly mimic those shown in Fig. \ref{fig:graph}. The
solid, uncorrected curve is extremely sensitive to errors in the
strength of the interaction, and therefore also to the exact placement
of the donor. The fidelity of the composite pulse follows the first
dotted line in Fig. \ref{fig:graph}. This curve is much less sensitive
to errors. As noted above, this pulse improves over the naive case for
$\Delta=-1$ to $\Delta=1$.


We now consider a largely random coupling between qubits. Remarkably,
regardless of our incomplete knowledge of the system, in many cases,
we can still create a high fidelity CNOT gate. To demonstrate this, we
consider the effect of random systematic error on the fidelity of a
CNOT gate. We will assume that the interaction Hamiltonian is given by
\begin{eqnarray}
H_R &=& J\left(\X\X + \Y \Y + \Z\Z \right) \nonumber \\
  & & + R \sum_{i,j = \{I,X,Y,Z\}} J_{ij}^{r} \sigma_i \sigma_j.
\end{eqnarray}
The coefficients $J_{ij}^{r}$ are chosen uniformly at random between
-1 and 1. The factor $R$ gives the strength of the random term in the
Hamiltonian.

The first three terms in this Hamiltonian give a simple Heisenberg
interaction. We did not need to choose the Heisenberg interaction. Any
coupling can be chosen, for example an Ising interaction, a
dipole-dipole interaction, or an XY interaction. Each different
coupling requires only trivial modifications. The non-random term in
the Hamiltonian represents the interaction or combination of
interactions expected to be present in the quantum system. The random
term in the Hamiltonian contributes to both single qubit terms, and
two qubit terms. The random term models our incomplete knowledge, not
only of the strength of the interaction, but also of the form of the
interaction.

\begin{figure}
\begin{center}
\includegraphics[angle=-90, width=\columnwidth]{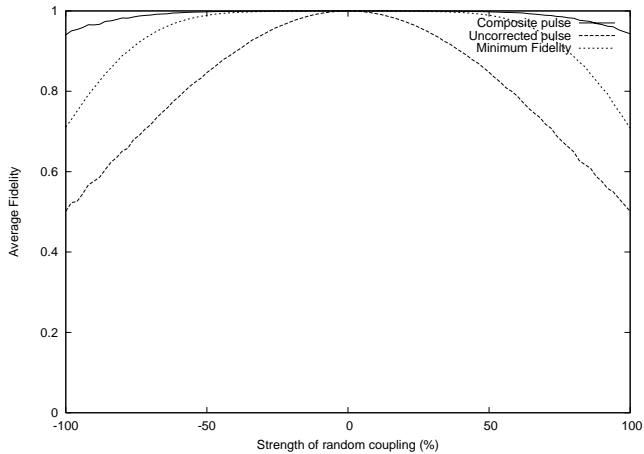}
\caption{Graph showing the fidelity of uncorrected and composite
  pulses to a Hamiltonian with a random component.} \label{fig:random}
\end{center}
\end{figure}

This demonstration uses the well known `square root of swap'
construction of the CNOT gate \cite{LD98}. Fig. \ref{fig:random} shows
the average and minimum fidelity of this construction, for different
values of $R/J$. The fidelity for each value of $R/J$ is calculated
for 1000 different random Hamiltonians. When the random contribution
of the Hamiltonian is large ($R/J = \pm 1$), the uncorrected CNOT gate
is useless. It has a average fidelity of approximately $50 \%$. This
is worse than if no interaction had been applied at all. Even in the
worst case of minimum fidelity, the composite pulse has a fidelity
superior to the uncorrected case.

If the `square root of swap' construction is replaced by the
composite pulse described in this paper, on average, a high fidelity
CNOT gate may be constructed. The mean fidelity, averaged over 1000
different random Hamiltonians for each value of $R/J$, is shown as
the dotted line in Fig. \ref{fig:random}. We also found the minimum
fidelity for the composite pulse, and this is also plotted in Fig.
\ref{fig:random}. Even when the random contribution is as large as
the exchange coupling, $R/J =1$, the average fidelity of the
composite gate is approximately $95\%$.

To obtain this composite pulse scheme, we combined decoupling with a
composite pulse scheme. First, the pulse used in Fig.
\ref{fig:random} was obtained using Eq. (\ref{eqn:refocus1}). Term
isolation was applied using $k=20$ repetitions for each gate.
Second, Eq. \ref{eqn:comp2} was used to correct the strength of the
$\Z\Z$ term. As Fig. \ref{fig:random} shows, a large increase in
fidelity is obtained. Even when the error is large and the coupling
between the qubits is essentially random, using the pulse schemes
presented in this paper, it is possible to produce a high fidelity
CNOT gate.


We have presented a method for creating CNOT gates which corrects
for systematic errors in both the form and strength of the
interaction between qubits. We have applied the composite pulse to a
model electron spin architecture, showing it does not slow down the
gates times too much to still fall below the fault tolerant
threshold. We also considered random systematic errors, showing that
the form of the systematic error is not important. The pulse scheme
presented here has broad applicability. Any system which implements
single qubit operations, and has a direct coupling between two
qubits directly may implement the composite pulses presented here.
In this paper we have shown that, regardless of an incomplete
knowledge of the strength or form of the interaction between two
qubits, in many cases it is possible to construct a CNOT gate which
has arbitrarily high fidelity.


I wish to thank Gerard Milburn, Hsi-Sheng Goan and Lloyd Hollenberg
for discussions and support in preparing this paper.

\bibliography{bibliography}

\begin{thebibliography}{21}
\expandafter\ifx\csname natexlab\endcsname\relax\def\natexlab#1{#1}\fi
\expandafter\ifx\csname bibnamefont\endcsname\relax
  \def\bibnamefont#1{#1}\fi
\expandafter\ifx\csname bibfnamefont\endcsname\relax
  \def\bibfnamefont#1{#1}\fi
\expandafter\ifx\csname citenamefont\endcsname\relax
  \def\citenamefont#1{#1}\fi
\expandafter\ifx\csname url\endcsname\relax
  \def\url#1{\texttt{#1}}\fi
\expandafter\ifx\csname urlprefix\endcsname\relax\def\urlprefix{URL }\fi
\providecommand{\bibinfo}[2]{#2}
\providecommand{\eprint}[2][]{\url{#2}}

\bibitem[{\citenamefont{Koiller et~al.}(2003)\citenamefont{Koiller, Hu, Drew,
  and Sarma}}]{KHD+03}
\bibinfo{author}{\bibfnamefont{B.}~\bibnamefont{Koiller}},
  \bibinfo{author}{\bibfnamefont{X.}~\bibnamefont{Hu}},
  \bibinfo{author}{\bibfnamefont{H.~D.} \bibnamefont{Drew}}, \bibnamefont{and}
  \bibinfo{author}{\bibfnamefont{S.~D.} \bibnamefont{Sarma}},
  \bibinfo{journal}{Physical Review Letters} \textbf{\bibinfo{volume}{90}},
  \bibinfo{eid}{067401} (pages~\bibinfo{numpages}{4}) (\bibinfo{year}{2003}).

\bibitem[{\citenamefont{Koiller
  et~al.}(2002{\natexlab{a}})\citenamefont{Koiller, Hu, and Sarma}}]{KHS02}
\bibinfo{author}{\bibfnamefont{B.}~\bibnamefont{Koiller}},
  \bibinfo{author}{\bibfnamefont{X.}~\bibnamefont{Hu}}, \bibnamefont{and}
  \bibinfo{author}{\bibfnamefont{S.~D.} \bibnamefont{Sarma}},
  \bibinfo{journal}{Phys. Rev. B} \textbf{\bibinfo{volume}{66}},
  \bibinfo{eid}{115201} (pages~\bibinfo{numpages}{12})
  (\bibinfo{year}{2002}{\natexlab{a}}).

\bibitem[{\citenamefont{Wellard et~al.}(2003)\citenamefont{Wellard, Hollenberg,
  Parisoli, Kettle, Goan, McIntosh, and Jamieson}}]{WHP+03}
\bibinfo{author}{\bibfnamefont{C.~J.} \bibnamefont{Wellard}},
  \bibinfo{author}{\bibfnamefont{L.~C.~L.} \bibnamefont{Hollenberg}},
  \bibinfo{author}{\bibfnamefont{F.}~\bibnamefont{Parisoli}},
  \bibinfo{author}{\bibfnamefont{L.~M.} \bibnamefont{Kettle}},
  \bibinfo{author}{\bibfnamefont{H.-S.} \bibnamefont{Goan}},
  \bibinfo{author}{\bibfnamefont{J.~A.~L.} \bibnamefont{McIntosh}},
  \bibnamefont{and} \bibinfo{author}{\bibfnamefont{D.~N.}
  \bibnamefont{Jamieson}}, \bibinfo{journal}{Phys. Rev. B}
  \textbf{\bibinfo{volume}{68}}, \bibinfo{pages}{195209}
  (\bibinfo{year}{2003}).

\bibitem[{\citenamefont{Wellard et~al.}(2004)\citenamefont{Wellard, Hollenberg,
  Kettle, and Goan}}]{WHK+04}
\bibinfo{author}{\bibfnamefont{C.~J.} \bibnamefont{Wellard}},
  \bibinfo{author}{\bibfnamefont{L.~C.~L.} \bibnamefont{Hollenberg}},
  \bibinfo{author}{\bibfnamefont{L.~M.} \bibnamefont{Kettle}},
  \bibnamefont{and} \bibinfo{author}{\bibfnamefont{H.-S.} \bibnamefont{Goan}},
  \bibinfo{journal}{J. Phys.: Condens. Matter} \textbf{\bibinfo{volume}{16}},
  \bibinfo{pages}{5697} (\bibinfo{year}{2004}).

\bibitem[{\citenamefont{Wellard and Hollenberg}(2005)}]{WH05}
\bibinfo{author}{\bibfnamefont{C.~J.} \bibnamefont{Wellard}} \bibnamefont{and}
  \bibinfo{author}{\bibfnamefont{L.~C.~L.} \bibnamefont{Hollenberg}},
  \bibinfo{journal}{Micro- and Nanotechnology: Materials, Processes, Packaging,
  and Systems II} \textbf{\bibinfo{volume}{5650}}, \bibinfo{pages}{94}
  (\bibinfo{year}{2005}).

\bibitem[{\citenamefont{Kettle et~al.}(2006)\citenamefont{Kettle, Goan, and
  Smith}}]{Ket06}
\bibinfo{author}{\bibfnamefont{L.~M.} \bibnamefont{Kettle}},
  \bibinfo{author}{\bibfnamefont{H.-S.} \bibnamefont{Goan}}, \bibnamefont{and}
  \bibinfo{author}{\bibfnamefont{S.~C.} \bibnamefont{Smith}},
  \bibinfo{journal}{Phys. Rev. B} \textbf{\bibinfo{volume}{73}},
  \bibinfo{pages}{115205} (\bibinfo{year}{2006}).

\bibitem[{\citenamefont{Freeman}(1997)}]{Fre97}
\bibinfo{author}{\bibfnamefont{R.}~\bibnamefont{Freeman}},
  \emph{\bibinfo{title}{Spin {C}horeography}} (\bibinfo{publisher}{Spektrum},
  \bibinfo{address}{Oxford}, \bibinfo{year}{1997}).

\bibitem[{\citenamefont{Levitt}(1986)}]{Lev86}
\bibinfo{author}{\bibfnamefont{M.~H.} \bibnamefont{Levitt}},
  \bibinfo{journal}{Progress in Nuclear Magnetic Resonance Spectroscopy}
  \textbf{\bibinfo{volume}{18}}, \bibinfo{pages}{61} (\bibinfo{year}{1986}).

\bibitem[{\citenamefont{Jones}(2003)}]{Jon03}
\bibinfo{author}{\bibfnamefont{J.~A.} \bibnamefont{Jones}},
  \bibinfo{journal}{Phys. Rev. A} \textbf{\bibinfo{volume}{67}},
  \bibinfo{eid}{012317} (pages~\bibinfo{numpages}{3}) (\bibinfo{year}{2003}).

\bibitem[{\citenamefont{Wimperis}(1984)}]{Wim84}
\bibinfo{author}{\bibfnamefont{S.}~\bibnamefont{Wimperis}},
  \bibinfo{journal}{J. Magn. Reson., Ser. B} \textbf{\bibinfo{volume}{109}},
  \bibinfo{pages}{221} (\bibinfo{year}{1984}).

\bibitem[{\citenamefont{Riebe et~al.}(2004)\citenamefont{Riebe, Haeffner, Roos,
  Haensel, Benhelm, Lancaster, Koerber, Becher, Schmidt-Kaler, James
  et~al.}}]{RHR+04}
\bibinfo{author}{\bibfnamefont{M.}~\bibnamefont{Riebe}},
  \bibinfo{author}{\bibfnamefont{H.}~\bibnamefont{Haeffner}},
  \bibinfo{author}{\bibfnamefont{C.~F.} \bibnamefont{Roos}},
  \bibinfo{author}{\bibfnamefont{W.}~\bibnamefont{Haensel}},
  \bibinfo{author}{\bibfnamefont{J.}~\bibnamefont{Benhelm}},
  \bibinfo{author}{\bibfnamefont{G.~P.~T.} \bibnamefont{Lancaster}},
  \bibinfo{author}{\bibfnamefont{T.~W.} \bibnamefont{Koerber}},
  \bibinfo{author}{\bibfnamefont{C.}~\bibnamefont{Becher}},
  \bibinfo{author}{\bibfnamefont{F.}~\bibnamefont{Schmidt-Kaler}},
  \bibinfo{author}{\bibfnamefont{D.~F.~V.} \bibnamefont{James}},
  \bibnamefont{et~al.}, \bibinfo{journal}{Nature}
  \textbf{\bibinfo{volume}{429}}, \bibinfo{pages}{734} (\bibinfo{year}{2004}).

\bibitem[{\citenamefont{Barrett et~al.}(2004)\citenamefont{Barrett, Chiaverini,
  Schaetz, Britton, Itano, Jost, Knill, Langer, Leibfried, Ozeri
  et~al.}}]{BCS+04}
\bibinfo{author}{\bibfnamefont{M.~D.} \bibnamefont{Barrett}},
  \bibinfo{author}{\bibfnamefont{J.}~\bibnamefont{Chiaverini}},
  \bibinfo{author}{\bibfnamefont{T.}~\bibnamefont{Schaetz}},
  \bibinfo{author}{\bibfnamefont{J.}~\bibnamefont{Britton}},
  \bibinfo{author}{\bibfnamefont{W.~M.} \bibnamefont{Itano}},
  \bibinfo{author}{\bibfnamefont{J.~D.} \bibnamefont{Jost}},
  \bibinfo{author}{\bibfnamefont{E.}~\bibnamefont{Knill}},
  \bibinfo{author}{\bibfnamefont{C.}~\bibnamefont{Langer}},
  \bibinfo{author}{\bibfnamefont{D.}~\bibnamefont{Leibfried}},
  \bibinfo{author}{\bibfnamefont{R.}~\bibnamefont{Ozeri}},
  \bibnamefont{et~al.}, \bibinfo{journal}{Nature}
  \textbf{\bibinfo{volume}{429}}, \bibinfo{pages}{737} (\bibinfo{year}{2004}).

\bibitem[{\citenamefont{Gulde et~al.}(2004)\citenamefont{Gulde, Riebe,
  Lancaster, Becher, Eschner, Haeffner, Schmidt-Kaler, Chuang, and
  Blatt}}]{GRL+04}
\bibinfo{author}{\bibfnamefont{S.}~\bibnamefont{Gulde}},
  \bibinfo{author}{\bibfnamefont{M.}~\bibnamefont{Riebe}},
  \bibinfo{author}{\bibfnamefont{G.~P.~T.} \bibnamefont{Lancaster}},
  \bibinfo{author}{\bibfnamefont{C.}~\bibnamefont{Becher}},
  \bibinfo{author}{\bibfnamefont{J.}~\bibnamefont{Eschner}},
  \bibinfo{author}{\bibfnamefont{H.}~\bibnamefont{Haeffner}},
  \bibinfo{author}{\bibfnamefont{F.}~\bibnamefont{Schmidt-Kaler}},
  \bibinfo{author}{\bibfnamefont{I.~L.} \bibnamefont{Chuang}},
  \bibnamefont{and} \bibinfo{author}{\bibfnamefont{R.}~\bibnamefont{Blatt}},
  \bibinfo{journal}{Nature} \textbf{\bibinfo{volume}{421}}, \bibinfo{pages}{48}
  (\bibinfo{year}{2004}).

\bibitem[{\citenamefont{Collin et~al.}(2004)\citenamefont{Collin, Ithier,
  Aassime, Joyez, Vion, and Esteve}}]{CIA+04}
\bibinfo{author}{\bibfnamefont{E.}~\bibnamefont{Collin}},
  \bibinfo{author}{\bibfnamefont{G.}~\bibnamefont{Ithier}},
  \bibinfo{author}{\bibfnamefont{A.}~\bibnamefont{Aassime}},
  \bibinfo{author}{\bibfnamefont{P.}~\bibnamefont{Joyez}},
  \bibinfo{author}{\bibfnamefont{D.}~\bibnamefont{Vion}}, \bibnamefont{and}
  \bibinfo{author}{\bibfnamefont{D.}~\bibnamefont{Esteve}}
  (\bibinfo{year}{2004}), \eprint{cond-mat/0404503}.

\bibitem[{\citenamefont{Kane}(1998)}]{Kan98}
\bibinfo{author}{\bibfnamefont{B.~E.} \bibnamefont{Kane}},
  \bibinfo{journal}{Nature} \textbf{\bibinfo{volume}{393}},
  \bibinfo{pages}{133} (\bibinfo{year}{1998}).

\bibitem[{\citenamefont{Bremner et~al.}(2004)\citenamefont{Bremner, Dodd,
  Nielsen, and Bacon}}]{BDN+04}
\bibinfo{author}{\bibfnamefont{M.~J.} \bibnamefont{Bremner}},
  \bibinfo{author}{\bibfnamefont{J.~L.} \bibnamefont{Dodd}},
  \bibinfo{author}{\bibfnamefont{M.~A.} \bibnamefont{Nielsen}},
  \bibnamefont{and} \bibinfo{author}{\bibfnamefont{D.}~\bibnamefont{Bacon}},
  \bibinfo{journal}{Phys. Rev. A} \textbf{\bibinfo{volume}{69}},
  \bibinfo{eid}{012313} (pages~\bibinfo{numpages}{12}) (\bibinfo{year}{2004}).

\bibitem[{\citenamefont{Brown et~al.}(2004)\citenamefont{Brown, Harrow, and
  Chuang}}]{BHC04}
\bibinfo{author}{\bibfnamefont{K.}~\bibnamefont{Brown}},
  \bibinfo{author}{\bibfnamefont{A.}~\bibnamefont{Harrow}}, \bibnamefont{and}
  \bibinfo{author}{\bibfnamefont{I.}~\bibnamefont{Chuang}},
  \bibinfo{journal}{Phys. Rev. A} \textbf{\bibinfo{volume}{70}},
  \bibinfo{pages}{052318} (\bibinfo{year}{2004}).

\bibitem[{\citenamefont{Koiller
  et~al.}(2002{\natexlab{b}})\citenamefont{Koiller, Hu, and
  Das~Sarma}}]{KHDS02}
\bibinfo{author}{\bibfnamefont{B.}~\bibnamefont{Koiller}},
  \bibinfo{author}{\bibfnamefont{X.}~\bibnamefont{Hu}}, \bibnamefont{and}
  \bibinfo{author}{\bibfnamefont{S.}~\bibnamefont{Das~Sarma}},
  \bibinfo{journal}{Phys. Rev. Lett.} \textbf{\bibinfo{volume}{88}},
  \bibinfo{pages}{027903} (\bibinfo{year}{2002}{\natexlab{b}}).

\bibitem[{\citenamefont{Hill et~al.}(2005)\citenamefont{Hill, Hollenberg,
  Fowler, Wellard, Greentree, and Goan}}]{Hil05}
\bibinfo{author}{\bibfnamefont{C.~D.} \bibnamefont{Hill}},
  \bibinfo{author}{\bibfnamefont{L.~C.~L.} \bibnamefont{Hollenberg}},
  \bibinfo{author}{\bibfnamefont{A.~G.} \bibnamefont{Fowler}},
  \bibinfo{author}{\bibfnamefont{C.~J.} \bibnamefont{Wellard}},
  \bibinfo{author}{\bibfnamefont{A.~D.} \bibnamefont{Greentree}},
  \bibnamefont{and} \bibinfo{author}{\bibfnamefont{H.-S.} \bibnamefont{Goan}},
  \bibinfo{journal}{Phys. Rev. B} \textbf{\bibinfo{volume}{72}},
  \bibinfo{pages}{045350} (\bibinfo{year}{2005}).

\bibitem[{\citenamefont{Gottesman}(1997)}]{Got97}
\bibinfo{author}{\bibfnamefont{D.}~\bibnamefont{Gottesman}}, Ph.D. thesis,
  \bibinfo{school}{California Institute of Technology} (\bibinfo{year}{1997}),
  \eprint{quant-ph/9705052}.

\bibitem[{\citenamefont{Loss and DiVincenzo}(1998)}]{LD98}
\bibinfo{author}{\bibfnamefont{D.}~\bibnamefont{Loss}} \bibnamefont{and}
  \bibinfo{author}{\bibfnamefont{D.~P.} \bibnamefont{DiVincenzo}},
  \bibinfo{journal}{Phys. Rev. A} \textbf{\bibinfo{volume}{57}},
  \bibinfo{pages}{120} (\bibinfo{year}{1998}).

\end{thebibliography}

\end{document}